\documentclass[paper]{ieice}
\usepackage[dvips]{graphicx}
\usepackage[fleqn]{amsmath}
\usepackage[varg]{txfonts}
\usepackage[psamsfonts]{amssymb}
\usepackage{balance}
\usepackage{cite}
\usepackage{latexsym}

\newtheorem{remark}{Remark}
\newcommand{\argmax}{\mathop{\mathrm{argmax}}\limits} 

\field{A}
\title[A Greedy Algorithm of Data-Dependent User Selection for Fast Fading Gaussian VBCs]{A Greedy Algorithm of Data-Dependent User Selection for Fast Fading Gaussian Vector Broadcast Channels}
\titlenote{ 
The work of K.~Takeuchi was in part supported by the Grant-in-Aid for 
Young Scientists~(B) (No.~23760329) from MEXT, Japan.}
\authorlist{
 \authorentry[ktakeuchi@uec.ac.jp]{Keigo TAKEUCHI}{m}{UEC}
 \authorentry[kawabata@uec.ac.jp]{Tsutomu KAWABATA}{m}{UEC}
}
\affiliate[UEC]{K.~Takeuchi and T.~Kawabata are with the Department of Communication Engineering and Informatics, the University of Electro-Communications, 
Tokyo 182-8585, Japan.}

\received{2011}{1}{1}
\revised{2011}{1}{1}

\begin{document}
\maketitle
\begin{summary}
User selection (US) with Zero-forcing beamforming is considered in 
fast fading Gaussian vector broadcast channels with perfect channel 
state information (CSI) at the transmitter. A novel criterion for US is 
proposed, which depends on both CSI and the data 
symbols, while conventional criteria only depend on CSI. Since the 
optimization of US based on the proposed criterion is infeasible, a greedy 
algorithm of data-dependent US is proposed to perform the optimization 
approximately. An overhead issue arises in fast fading channels: On every 
update of US, the transmitter might inform each user whether he/she has been 
selected, using a certain fraction of resources. This overhead results in a 
significant rate loss for fast fading channels. In order to circumvent this 
overhead issue, iterative detection and decoding schemes are proposed on the 
basis of belief propagation. The proposed iterative schemes require no 
information about whether each user has been selected. The proposed US scheme 
is compared to a data-independent US scheme. The complexity of the two schemes 
is comparable to each other for fast fading channels. Numerical simulations 
show that the proposed scheme can outperform the data-independent scheme 
for fast fading channels in terms of energy efficiency, bit error rate, and 
achievable sum rate. 
\end{summary}

\section{Introduction} \label{sec1} 
Multiple-input multiple-output broadcast channels (MIMO-BCs) 
are a mathematical model of downlink channels in which a base station with 
multiple transmit antennas communicates with multiple receivers (users). In 
this paper, they are referred to as vector broadcast channels (VBCs) since the 
number of receive antennas is assumed to be one.   

The capacity region of MIMO-BCs has been shown to be achieved by 
dirty-paper coding (DPC)~\cite{Caire03,Viswanath03,Yu04,Weingarten06}. 
DPC~\cite{Costa83} is a sophisticated precoding scheme that pre-cancels 
multiple-access interference (MAI) to each user at the transmitter, by 
utilizing the information about the data symbols transmitted to the other 
users. However, DPC is infeasible because of the high 
complexity. Thus, a recent research issue is to construct a precoding scheme 
that can achieve an appropriate tradeoff between the complexity and 
the performance.   

In order to achieve a good tradeoff between the complexity and the performance, 
user selection (US) with zero-forcing (ZF) beamforming (ZFBF) 
has been considered~\cite{Tu03,Dimic05,Yoo06}. 
US is based on a different idea from that for DPC: US aims to keep the MAI 
power as small as possible by selecting a subset of channel vectors with 
higher orthogonality. On the other hand, DPC pre-cancels (possibly large) 
MAI by utilizing the information about the data symbols as well as channel 
state information (CSI). 
Yoo and Goldsmith~\cite{Yoo06} proved that a greedy algorithm of US with 
ZFBF can achieve the sum capacity as the number of users tends to infinity, 
even though it utilizes no information about the data symbols. Intuitively, 
this result can be understood as follows: The algorithm attempts to select a 
subset of channel vectors with higher orthogonality. When the number of users 
tends to infinity, it is possible to select a subset of users whose channel 
vectors are almost orthogonal. Consequently, the algorithm can achieve the sum 
capacity in that limit, while it is suboptimal for a finite number of users. 

A crucial assumption for US is the assumption of quasi-static or show fading 
channels. This assumption becomes unrealistic as {\em mobility} of users 
increases. Thus, it is important in practice to investigate fast fading 
channels. Note that the meaning of fast or slow is relative. In this paper, 
fading is said to be fast when the coherence time is much shorter than the 
code length, determined by delay constraints~\cite{Tse05}. 

The purpose of this paper is to construct a novel US-based communication 
scheme that is suitable for fast fading channels. An overhead issue arises 
in fast fading channels: US should be updated frequently for fast fading 
channels to keep track of the fading channels. On every update of US, the base 
station might inform each user whether he/she has been selected, using a 
certain fraction of resources. This overhead is negligibly small for 
quasi-static or slow fading channels, since the frequency of updates is low. 
However, the frequency of updates grows as the coherence time of fading 
channels reduces. Consequently, the overhead results in a large rate loss for 
fast fading channels. In order to circumvent this overhead issue, we propose a 
communication scheme that allows each user to detect whether he/she has been 
selected with no overhead.  

It is possible to attain an additional gain in performance for fast fading 
channels. We propose a {\em data-dependent} criterion of US that combines the 
ideas of US and DPC, while the existing criteria of conventional US are 
data-independent~\cite{Tu03,Dimic05,Yoo06}. A greedy algorithm of 
data-dependent US based on the proposed criterion is systematically derived to 
select a subset of channel vectors with high orthogonality {\em and} to 
pre-cancel MAI by using the information about the data symbols. 
A frequent update of US or a small block size of US results in decreasing 
the number of interfering signals that should be pre-cancelled simultaneously. 
Thus, MAI can be pre-cancelled well if the block size of US is small, or 
if US is updated frequently. In other words, the data-dependent US that 
pre-cancels MAI is suitable for fast fading channels. 

The rest of this paper is organized as follows: After summarizing the notation 
used in this paper, a VBC is introduced in Section~\ref{sec2} 
as a mathematical model of downlink channels. In Section~\ref{sec3} 
a novel criterion of US is proposed on the basis of a lower bound of the 
achievable sum rate for the fast fading VBC. Furthermore, we present a 
systematical derivation for a greedy algorithm of data-dependent US based on 
the proposed criterion. In Section~\ref{sec4} we propose iterative receivers 
that allow each user to detect whether he/she has been selected. Numerical 
simulations presented in Section~\ref{sec5} show that the data-dependent 
scheme can outperform data-independent US for fast fading channels. 
Section~\ref{sec6} concludes this paper. 

\subsection{Notation} 
Throughout this paper, $\vec{\boldsymbol{a}}$ denotes a row vector, while 
$\boldsymbol{a}$ represents a column vector. 
For a matrix $\boldsymbol{A}$, $\boldsymbol{A}^{\mathrm{T}}$ and 
$\boldsymbol{A}^{\mathrm{H}}$ stand for the transpose and the conjugate 
transpose of $\boldsymbol{A}$, respectively. For a full-rank 
matrix $\boldsymbol{A}\in\mathbb{C}^{K\times N}$, with $K\leq N$, 
$\boldsymbol{A}^{\dagger}$ denotes a pseudo-inverse of $\boldsymbol{A}$, 
given by  
\begin{equation}
\boldsymbol{A}^{\dagger}
= \boldsymbol{A}^{\mathrm{H}}
(\boldsymbol{A}\boldsymbol{A}^{\mathrm{H}})^{-1}. 
\end{equation}
The matrix $\boldsymbol{I}_{N}$ represents the $N$-dimensional identity 
matrix. A circularly symmetric complex Gaussian distribution with variance 
$\sigma^{2}$ is denoted by $\mathcal{CN}(0,\sigma^{2})$. For functions 
$f(x)$ and $g(x)$, $f(x)\propto g(x)$ means that $f(x)$ is proportional 
to $g(x)$, i.e., there is such a constant $C$ that $f(x)=Cg(x)$. 

\section{Channel Model} \label{sec2} 
We consider a $K$-user Gaussian VBC in which the base station has $N$ transmit 
antennas. The base station communicates with the users over $T$ time slots. 
The received signal $y_{k,t}\in\mathbb{C}$ of user~$k$ with one 
receive antenna in time slot~$t$ ($t=0,1,\ldots,T-1$) is given by 
\begin{equation} \label{VBC} 
y_{k,t} = \frac{1}{\sqrt{\mathcal{E}}}
\vec{\boldsymbol{h}}_{k,t}\boldsymbol{u}_{t} + n_{k,t}, 
\quad \hbox{for $k=1,\ldots,K$,}
\end{equation}
with 
\begin{equation} \label{energy_penalty} 
\mathcal{E}=\frac{1}{T}\sum_{t=0}^{T-1}\|\boldsymbol{u}_{t}\|^{2}. 
\end{equation}
In (\ref{VBC}), $\boldsymbol{u}_{t}\in\mathbb{C}^{N}$ and $n_{k,t}\sim
\mathcal{CN}(0,N_{0})$ denote the transmitted vector in time slot~$t$ and 
the additive white Gaussian noise (AWGN) with variance $N_{0}$ for 
user~$k$ in time slot~$t$, respectively. The row vector 
$\vec{\boldsymbol{h}}_{k,t}\in\mathbb{C}^{1\times N}$ represents the 
channel gains between the transmitter and user~$k$ with 
$\mathbb{E}[\vec{\boldsymbol{h}}_{k,t}^{\mathrm{H}}
\vec{\boldsymbol{h}}_{k,t}]=N^{-1}\boldsymbol{I}_{N}$. The assumption  
$\mathbb{E}[|(\vec{\boldsymbol{h}}_{k,t})_{n}|^{2}]=1/N$ 
normalizes the power gain obtained by increasing the number of transmit 
antennas. The coefficient $1/\sqrt{\mathcal{E}}$ in (\ref{VBC}) implies that the 
average transmit power is restricted to $1$. 

The over-loaded case $K\geq N$ is considered in this paper. 
The channel vectors $\{\vec{\boldsymbol{h}}_{k,t}:\hbox{for all $k$}\}$ 
for different users are assumed to be mutually independent. 
For simplicity, we assume perfect CSI at the transmitter, i.e., that all 
channel vectors $\{\vec{\boldsymbol{h}}_{k,t}\}$ are known to the 
transmitter. Note that the latter assumption is an idealized assumption for 
time-division duplex (TDD) systems. The influence of channel estimation errors 
will be briefly noted in Section~\ref{sec4}. For further simplifications, 
phase shift keying (PSK) is assumed, and power allocation is not considered 
in this paper. 

\begin{figure}[t]
\begin{center}
\includegraphics[width=\hsize]{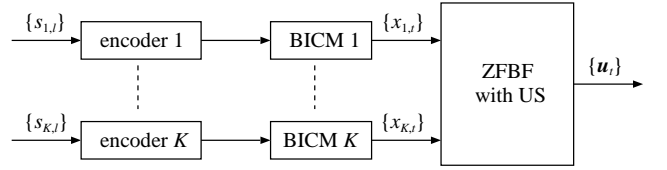}
\caption{
Transmitter. 
}
\label{fig1} 
\end{center}
\end{figure}

\section{Transmitter} \label{sec3} 
\subsection{Overview}
Figure~\ref{fig1} shows a diagram of the proposed transmitter. 
A binary information sequence $\{s_{k,l}\in\{0,1\}\}_{l=1}^{L}$ of 
length $L$ for user~$k$ is first encoded by a per-user encoder with  
rate~$r$. In order to combat burst errors, bit-interleaved coded modulation 
(BICM) with a PSK constellation $\mathcal{M}\subset\mathbb{C}$ is applied 
to the coded sequence. The obtained sequence 
$\{x_{k,t}\in\mathcal{M}:t=0,\ldots,T-1\}$, with $T=L/(r\log_{2}|\mathcal{M}|)$, 
is fed to a ZF beamformer with US, proposed in Section~\ref{sec_33}. 
Since power allocation is not considered, $\mathbb{E}[|x_{k,t}|^{2}]=1$ is 
assumed for all users. Let $\tilde{K}$ ($\leq\mathrm{min}\{K,N\}$) and $B$ 
denote the number of selected users and the block size of US, respectively. 
US is updated for every $B$ time slots, i.e., $\tilde{K}$ users are selected 
and fixed during $B$ time slots. Let $\mathcal{K}_{t}\subset\{1,\ldots,K\}$ 
denote the set of users selected in time slot~$t$. Note that 
$\{\mathcal{K}_{t}\}$ are the same for time slots belonging to the identical 
block of US. The vector $\boldsymbol{u}_{t}$ transmitted in time slot~$t$ is 
given by $\boldsymbol{u}_{t}=\boldsymbol{u}_{\mathcal{K}_{t},t}$~\cite{Wiesel08}, 
with 
\begin{equation} \label{ZF} 
\boldsymbol{u}_{\mathcal{K}_{t},t} = 
\boldsymbol{H}_{\mathcal{K}_{t},t}^{\dagger}
\boldsymbol{x}_{\mathcal{K}_{t},t}. 
\end{equation} 
In (\ref{ZF}), the matrix 
$\boldsymbol{H}_{\mathcal{K}_{t},t}\in\mathbb{C}^{\tilde{K}\times N}$ 
and the vector $\boldsymbol{x}_{\mathcal{K}_{t},t}\in\mathbb{C}^{\tilde{K}}$ 
are generated by stacking the channel vectors 
$\{\vec{\boldsymbol{h}}_{k,t}\in\mathbb{C}^{1\times N}:
k\in\mathcal{K}_{t}\}$ 
and the data symbols $\{x_{k,t}:k\in\mathcal{K}_{t}\}$ 
for the users $\mathcal{K}_{t}$ selected in time slot~$t$, respectively. 
They must be stacked in the same order, otherwise the data symbols would be 
sent to unintended users. The data symbols 
$\{x_{k,t}:k\notin\mathcal{K}_{t}\}$ for the non-selected users in time 
slot~$t$ are {\em discarded} at the transmitter side. They are recovered at 
the receiver by utilizing redundancy of the error-correcting code. 
The computational complexity required in the receiver can be reduced by 
discarding the data symbols for the non-selected users. The details will be 
remarked in the next section. 
 
Fast fading channels are considered in this paper, as mentioned in 
Section~\ref{sec1}: The code length $L/r$ or the length of interleaving 
$T=L/(r\log_{2}|\mathcal{M}|)$ is assumed to be much longer than 
the coherence time of the fading channels. Note that the dominant factor of 
delay is not US but the error-correcting, since the block size $B$ of US is 
comparable to the coherence time.  

\subsection{Criterion for US} \label{sec_32} 
Sum rate and fairness should be taken into account as criteria for selecting 
users. For simplicity, however, we only consider a criterion based on the 
achievable sum rate, and propose a novel criterion for the fast fading 
VBC~(\ref{VBC}). 

$T/B$ updates of US are performed during $T$ time slots, since the block size 
of US is $B$. We focus on block~$j$ of US for $j=0,1,\ldots,T/B-1$. 
Applying the ZFBF~(\ref{ZF}) to the VBC~(\ref{VBC}) implies that, if user~$k$ 
has been selected in block~$j$, he/she receives the sum of the normalized 
data symbol $\mathcal{E}^{-1/2}x_{k,t}$ and the AWGN $n_{k,t}$ in the 
corresponding time slots $t=jB,\ldots,jB+B-1$. Otherwise, user~$k$ receives 
the sum of the AWGN $n_{k,t}$ and the interference 
$\mathcal{E}^{-1/2}\vec{\boldsymbol{h}}_{k}\boldsymbol{u}_{\mathcal{K}_{t},t}$, 
with (\ref{ZF}), the later of which is caused by the ZFBF intended for the 
selected users. These observations imply that the equivalent channel for 
user~$k$ in block~$j$ is given by 
\begin{equation} \label{equivalent_channel} 
y_{k,t} = \frac{1}{\sqrt{\mathcal{E}}}\left\{
 a_{k}^{(j)}x_{k,t} + (1-a_{k}^{(j)})I_{k,t} 
\right\} + n_{k,t}, 
\end{equation}
for $t=jB,\ldots,jB+B-1$. In (\ref{equivalent_channel}), 
$I_{k,t}=\vec{\boldsymbol{h}}_{k}\boldsymbol{u}_{\mathcal{K}_{t},t}$ denotes 
the interference to user~$k$, with (\ref{ZF}). Furthermore, 
$a_{k}^{(j)}\in\{0,1\}$ indicates whether user~$k$ has been selected in 
block~$j$, i.e.\ 
\begin{equation} \label{a_k} 
a_{k}^{(j)} = \left\{
\begin{array}{cl}
1 & \hbox{when user~$k$ is selected in the $j$th block} \\ 
0 & \hbox{otherwise.}
\end{array}
\right.
\end{equation}
Note that $a_{k}^{(j)}$ is unknown to the receiver in advance. 

\begin{remark}
Let us discuss why the data symbols for non-selected users should be discarded 
at the transmitter side. The received signal~(\ref{equivalent_channel}) in 
time slot~$t$ contains only the data symbol $x_{k,t}$ in the same time slot, 
because the transmitter has discarded the data symbols for non-selected users. 
What would occur if the transmitter kept the data symbols for the non-selected 
users? The received signal in time slot~$t$ might not contain the data symbol 
in the same time slot. Thus, each user would have to detect the index of the 
data symbol sent in time slot~$t$. A simple method is to count how many times 
he/she has been selected. This method cannot yield the correct index unless 
all decisions of $a_{k}$ in the preceding blocks are correct. Consequently, 
serious error propagation would occur once (\ref{a_k}) is detected 
incorrectly. This argument implies that a complicated receiver would be 
required for detecting (\ref{a_k}) if the data symbols were not discarded. 
This is the reason why the data symbols for non-selected users should be 
discarded at the transmitter side. 
\end{remark}

We shall assess the achievable sum rate for user~$k$. 
Let us assume that (\ref{a_k}) can be detected 
with no errors. This assumption can be a reasonable assumption even for small 
$B$, as demonstrated numerically in Section~\ref{sec5}. In this case, 
the equivalent channel~(\ref{equivalent_channel}) can be regarded as a 
Gaussian erasure channel, in which each erasure probability 
$\mathrm{Prob}(a_{k}^{(j)}=0)$ may depend on the data symbol 
$x_{k,t}$ for user~$k$ via the set of selected users $\mathcal{K}_{t}$. 
We ignore this dependencies in this paper. The achievable rate under this 
assumption should provide a lower bound on the true one, since the receiver 
can obtain information about the data symbols from the observations of 
$\{a_{k}^{(j)}\}$. 
In order to evaluate the achievable rate, we need the average frequency at 
which each user is selected. The channel gains for each user become large or 
small block by block. This fading effect is averaged out for sufficiently 
large $T$, because of the assumption of fast fading. Consequently, the users 
should experience the identical channel quality in average, so that each user 
should be selected at a frequency of $\tilde{K}/K$ as $T\rightarrow\infty$. 
Since the effective signal-to-noise ratio (SNR) for each {\em selected} user 
is equal to $(\mathcal{E}N_{0})^{-1}$, from (\ref{equivalent_channel}), a lower 
bound $R_{k}$ on the achievable rate of user~$k$ for transmission over $T$ 
time slots is given by~\cite{Julian02}  
\begin{equation} \label{user_rate}
R_{k} = \frac{\tilde{K}}{K}C\left(
 \frac{1}{\bar{\mathcal{E}}N_{0}}
\right),  
\end{equation}
as $T\rightarrow\infty$, with
\begin{equation} \label{average_energy_penalty}
\bar{\mathcal{E}}=\lim_{T\rightarrow\infty}
\frac{1}{T}\sum_{t=0}^{T-1}\|\boldsymbol{u}_{t}\|^{2}. 
\end{equation}
In (\ref{user_rate}), $C(\gamma)$ denotes the achievable rate of the 
AWGN channel with the signal-to-noise ratio (SNR) $\gamma$, defined as the 
mutual information between the data symbol and the received signal 
$I(x_{k,t};y_{k,t}|a_{k}^{(j)}=1)$~\cite{Cover06}. See \ref{deriv_user_rate} for 
the formal derivation of (\ref{user_rate}). Equation~(\ref{user_rate}) implies 
that a lower bound $R$ on the achievable sum rate for the fast fading 
VBC~(\ref{VBC}) is given by 
\begin{equation} \label{sum_rate} 
R = \sum_{k=1}^{K}R_{k}=\tilde{K}C\left(
 \frac{1}{\bar{\mathcal{E}}N_{0}}
\right). 
\end{equation}

\begin{remark}
In the derivation of (\ref{user_rate}), we have implicitly assumed that 
data-dependent US does not change the distribution of the data 
symbol~$x_{k,t}$. This assumption is valid for PSK data symbols considered 
in this paper. However, the assumption does not hold for multi-level 
modulation, since the transmitter can reduce (\ref{average_energy_penalty}) 
by selecting users who transmit the data symbols with small amplitudes.   
\end{remark}

Maximizing the achievable sum rate~(\ref{sum_rate}) for given $\tilde{K}$ and 
$B$ is equivalent to minimizing (\ref{average_energy_penalty}), since the 
achievable rate $C(\gamma)$ is a monotonically increasing function of 
$\gamma$. This conclusion is due to the assumption of equal powers for all 
users, i.e.\ $\mathbb{E}[|x_{k,t}|^{2}]=1$. If power allocation were used, 
maximizing the achievable sum rate might not be equivalent to minimizing 
(\ref{average_energy_penalty}). The average 
power~(\ref{average_energy_penalty}) of the transmitted vector should not 
be confused with the average transmit power, which is restricted to $1$ 
owing to the coefficient $1/\sqrt{\mathcal{E}}$ in (\ref{VBC}). The average 
power~(\ref{average_energy_penalty}) should be regarded as a cost for 
performing ZFBF~(\ref{ZF}). Thus, we hereafter refer to 
(\ref{average_energy_penalty}) as {\em energy penalty}. 

We first minimize the energy penalty for given $\tilde{K}$ and $B$. 
The number of selected users $\tilde{K}$ is chosen so as to maximize the 
achievable sum rate~(\ref{sum_rate}). On the other hand, the block size 
of US $B$ should be selected carefully on the basis of the energy penalty and 
the detection performance for (\ref{a_k}). The details will be discussed in 
Section~\ref{sec53}. The minimum of (\ref{average_energy_penalty}) for fixed 
$\tilde{K}$ and $B$, denoted by $\bar{\mathcal{E}}_{\min}$, is achieved 
when the time average of $\{\|\boldsymbol{u}_{t}\|^{2}\}$ in each block is 
minimized:   
\begin{align}  
\bar{\mathcal{E}}_{\min} 
=& \lim_{T\rightarrow\infty}\frac{B}{T}\sum_{j=0}^{T/B-1}
\min_{\mathcal{K}\subset\{1,\ldots,K\}}E_{\mathcal{K}}^{(j)}(B)  
\nonumber \\ 
=& \mathbb{E}\left[
 \min_{\mathcal{K}\subset\{1,\ldots,K\}}E_{\mathcal{K}}^{(j)}(B)  
\right], \label{energy_min}
\end{align}
with the number of selected users fixed $|\mathcal{K}|=\tilde{K}$. 
In (\ref{energy_min}), $E_{\mathcal{K}}^{(j)}(B)$ denotes the energy penalty 
for block~$j$, 
\begin{equation} \label{instantaneous_power}
E_{\mathcal{K}}^{(j)}(B) = \frac{1}{B}\sum_{t=0}^{B-1}
\|\boldsymbol{u}_{\mathcal{K},t+jB}\|^{2}, 
\end{equation}
where $\boldsymbol{u}_{\mathcal{K},t}$ is defined as (\ref{ZF}). 

In conventional US, the energy penalty~(\ref{instantaneous_power}) 
may be minimized {\em after} taking the limit $B\rightarrow\infty$, in which 
(\ref{instantaneous_power}) converges in probability to the conditional 
expectation with respect to the data symbols, 
\begin{equation} \label{averaged_energy_penalty} 
\mathcal{E}_{\mathcal{K}}^{(j)} = \mathbb{E}\left[
 \left. 
  \|\boldsymbol{u}_{\mathcal{K},t+jB}\|^{2} 
 \right| \{\vec{\boldsymbol{h}}_{k,t}\} 
\right].
\end{equation}  
The minimizer of the energy penalty~(\ref{instantaneous_power}) depends on 
both channel vectors and data symbols, while the conventional criterion never 
depends on the realizations of data symbols. 
Note that the minimization and the limit $B\rightarrow\infty$ are not 
necessarily commutative. It is straightforward to find that the energy penalty 
based on data-dependent US is smaller than the conventional one: Let 
$\mathcal{K}_{\mathrm{con}}$ denote the minimizer of the 
energy penalty~(\ref{averaged_energy_penalty}) averaged over the data symbols. 
In both sides of the inequality 
\begin{equation}
\min_{\mathcal{K}}E_{\mathcal{K}}^{(j)}(B) 
\leq E_{\mathcal{K}_{\mathrm{con}}}^{(j)}(B), 
\end{equation}
we take the limit $B\rightarrow\infty$. Since $\mathcal{K}_{\mathrm{con}}$ 
is independent of the data symbols, we can use the weak law of large 
numbers to find that the right-hand side tends to the minimum of 
(\ref{averaged_energy_penalty}), i.e.\  
$\mathcal{E}_{\mathcal{K}_{\mathrm{cov}}}^{(j)}$. This implies  
\begin{equation}
\lim_{B\rightarrow\infty}\min_{\mathcal{K}}E_{\mathcal{K}}^{(j)}(B) 
\leq \min_{\mathcal{K}}\lim_{B\rightarrow\infty}
E_{\mathcal{K}}^{(j)}(B). 
\end{equation}

\subsection{Data-Dependent User Selection} \label{sec_33}
The minimization of the energy penalty~(\ref{instantaneous_power}) is 
infeasible, because of high complexity, as conventional criteria are. 
Instead, we propose a greedy algorithm to calculate the minimization 
approximately. 
Without loss of generality, we hereafter focus on the first $B$ time slots, 
and drop the superscript $^{(j)}$. A small value of the block size $B$ of US 
is used, so that one can postulate that the channels are 
fixed during one block of US, i.e.\ 
$\vec{\boldsymbol{h}}_{k,1}=\cdots=
\vec{\boldsymbol{h}}_{k,B}\equiv\vec{\boldsymbol{h}}_{k}$. 
For notational convenience, the set of users selected 
in time slots~$t=0,\ldots,B-1$ is denoted by $\mathcal{K}$. Furthermore,  
the matrix $\boldsymbol{H}_{\mathcal{K},t}$ 
is re-written as $\boldsymbol{H}_{\mathcal{K}}$.  

The derivation of the proposed algorithm is summarized in~\ref{appen1}. 
We first present several definitions used in the algorithm. 
In the proposed greedy algorithm users are selected one by one. 
Let $\mathcal{K}(i)\subset\{1,\ldots,K\}$ denote the set of users selected 
in the first $i$ stages, with $|\mathcal{K}(i)|=i$. 
The ZFBF vector $\boldsymbol{u}_{\mathcal{K}(i),t}\in\mathbb{C}^{N}$ for 
the users $\mathcal{K}(i)$ selected in the first $i$ stages is given by  
$\boldsymbol{u}_{\mathcal{K}(i),t}=\boldsymbol{H}_{\mathcal{K}(i)}^{\dagger} 
\boldsymbol{x}_{\mathcal{K}(i),t}$. Furthermore, 
$\boldsymbol{P}_{i}^{\perp}$ denotes the projection matrix from 
$\mathbb{C}^{1\times N}$ onto the orthogonal complement 
of the subspace spanned by the channel vectors 
$\{\vec{\boldsymbol{h}}_{k}:k\in\mathcal{K}(i)\}$ 
selected in the first $i$ stages. 
The two matrices $\boldsymbol{H}_{\mathcal{K}(i)}^{\dagger}$ 
and $\boldsymbol{P}_{i}^{\perp}$ are calculated recursively as follows: 
\begin{equation} \label{P} 
\boldsymbol{P}_{i}^{\perp} = \boldsymbol{I}_{N} 
- \boldsymbol{H}_{\mathcal{K}(i)}^{\dagger}\boldsymbol{H}_{\mathcal{K}(i)}, 
\end{equation}
\begin{equation} \label{H} 
\boldsymbol{H}_{\mathcal{K}(i)}^{\dagger} 
= \begin{bmatrix}
\left(
 \boldsymbol{I}_{N} - \frac{
  \boldsymbol{P}_{i-1}^{\perp}
  \vec{\boldsymbol{h}}_{\hat{k}}^{\mathrm{H}} 
  \vec{\boldsymbol{h}}_{\hat{k}}
 }{
  \vec{\boldsymbol{h}}_{\hat{k}}\boldsymbol{P}_{i-1}^{\perp}
  \vec{\boldsymbol{h}}_{\hat{k}}^{\mathrm{H}} 
 }
\right)\boldsymbol{H}_{\mathcal{K}(i-1)}^{\dagger}, & \frac{
 \boldsymbol{P}_{i-1}^{\perp}\vec{\boldsymbol{h}}_{\hat{k}}^{\mathrm{H}} 
}{\vec{\boldsymbol{h}}_{\hat{k}}\boldsymbol{P}_{i-1}^{\perp}
\vec{\boldsymbol{h}}_{\hat{k}}^{\mathrm{H}} }
\end{bmatrix}, 
\end{equation} 
where $\hat{k}$ denotes the user selected in stage~$i$. 

The proposed greedy algorithm selects the user $\hat{k}$ to minimize 
(\ref{instantaneous_power}) with $\mathcal{K}=\mathcal{K}(i)$   
in stage~$i$, which is recursively given by  
\begin{equation} \label{energy_rec} 
E_{\mathcal{K}(i)}(B) = \frac{
B^{-1}\sum_{t=1}^{B}|x_{k,t} - \vec{\boldsymbol{h}}_{k}
\boldsymbol{u}_{\mathcal{K}(i-1),t}|^{2}}
{\|\vec{\boldsymbol{h}}_{k}\boldsymbol{P}_{i-1}^{\perp}\|^{2}} 
+ E_{\mathcal{K}(i-1)}(B). 
\end{equation}
The proposed algorithm is summarized as follows: 

\begin{description}
\item[Step~1] $i=1$, $\mathcal{K}(0)=\emptyset$, 
$\boldsymbol{P}_{0}^{\perp}=\boldsymbol{I}_{N}$, 
$\boldsymbol{u}_{\emptyset,t}=\boldsymbol{0}$, and $E_{\emptyset}=0$.  
\item[Step~2] Let $\mathcal{K}(i)=\mathcal{K}(i-1)\cup\{\hat{k}\}$, 
where the user 
$\hat{k}\in\{1,\ldots,K\}\backslash\mathcal{K}(i-1)$ 
minimizes (\ref{energy_rec}). 
\item[Step~3] If $i=\tilde{K}$, outputs $\mathcal{K}=\mathcal{K}(i)$. 
Otherwise, compute (\ref{P}) and (\ref{H}), and go back to Step~2 after 
$i:=i+1$.  
\end{description}

Expression~(\ref{energy_rec}) provides a useful interpretation with respect to 
the proposed algorithm. In order to select the user minimizing 
(\ref{energy_rec}), one should select such a user that the denominator 
of the first term in the right-hand side of (\ref{energy_rec}) is large, or 
that the numerator is small. Existing US algorithms have been proposed 
on the basis of maximizing the denominator 
$\|\vec{\boldsymbol{h}}_{k}\boldsymbol{P}_{i-1}^{\perp}\|^{2}$~\cite{Tu03} or 
of its modifications~\cite{Dimic05,Yoo06}. 
Selecting the user to maximize the denominator is equivalent to selecting 
a user that achieves high orthogonality between his/her channel vector and 
the channel vectors selected in the preceding stages. 
The point of the proposed algorithm is that the numerator is also taken into 
account, along with the denominator. The numerator becomes small when the 
amplitude and phase of the interference 
$\vec{\boldsymbol{h}}_{k}\boldsymbol{u}_{\mathcal{K}(i-1),t}$ 
are close to those of the transmitted symbol $\boldsymbol{x}_{k,t}$. 
The proposed algorithm selects the user attaining an appropriate tradeoff 
between two criteria, i.e., between the maximization of the denominator and 
the minimization of the numerator. 

\begin{remark} \label{remark3} 
It is possible to derive a greedy algorithm of US based on the 
data-independent criterion~(\ref{averaged_energy_penalty}), instead of 
the data-dependent criterion~(\ref{instantaneous_power}). The obtained 
algorithm is equivalent to the restriction of a greedy algorithm proposed 
in \cite{Dimic05} to the case of no power allocation. We hereafter refer to 
this greedy algorithm as data-independent US. 
\end{remark}

Let us evaluate the computational complexity of the proposed algorithm. 
In Step~2 the computational costs for calculating (\ref{ZF}) with 
$\mathcal{K}_{t}=\mathcal{K}(i-1)$ and the numerator in (\ref{energy_rec}) 
are $O(iBN)$ and $O(BN)$  in stage~$i$, respectively. 
Furthermore, the complexity for evaluating the denominator in 
(\ref{energy_rec}) is given by $O(N^{2})$. Thus, the complexity required in 
Step~2 is $O\{iBN+K(BN+N^{2})\}$ in stage~$i$. 
Similarly, we find that the complexity needed in Step~3 
is $O(iN^{2})$ in stage~$i$. Thus, the complexity of the proposed algorithm 
is given by $O\{K\tilde{K}\max(B,N)N \}$, because of $\tilde{K}\leq\min(K,N)$. 

Decreasing the block size $B$ results in increasing the frequency of the US. 
Thus, we focus on the computational complexity per time slot, which is given by 
$O\{K\tilde{K}\max(1,N/B)N\}$ for the proposed algorithm. On the other hand, 
the complexity of the conventional US schemes proposed in 
\cite{Tu03,Dimic05,Yoo06}, including the data-independent scheme in 
Remark~\ref{remark3}, is $O(K\tilde{K}N^{2})$. Thus, the complexity per time 
slot is equal to $O\{K\tilde{K}(N/T_{\mathrm{c}})N)$, with $T_{\mathrm{c}}$ 
denoting the coherence time of fading channels, which is equal to the block 
size of the conventional US schemes. Imposing the constraint 
$B\leq T_{\mathrm{c}}$ implies $N/T_{\mathrm{c}}\leq\max(1,N/B)$, where the 
equality holds only when $N=B=T_{\mathrm{c}}$, because of 
$N/T_{\mathrm{c}}\leq N/B\leq \max(1,N/B)$. 
Thus, the complexity per time slot of the proposed scheme is the same as that 
for the conventional US schemes when $N=B=T_{\mathrm{c}}$. This result implies 
that the proposed algorithm is efficient in terms of the complexity for 
small $T_{\mathrm{c}}$, since it is not easy in practice to use many 
transmit antennas.  

\section{Iterative Receivers} \label{sec4} 
\subsection{Belief Propagation}  
\begin{figure}[t]
\begin{center}
\includegraphics[width=\hsize]{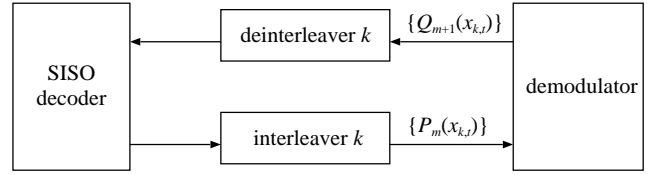}
\caption{
Iterative detection and decoding. 
}
\label{fig2} 
\end{center}
\end{figure}

The goal of the receiver is to perform the (bit-wise) maximum a posteriori 
(MAP) decoding of the information sequence $\{s_{k,l}\}$ given the received 
signals $\{y_{k,t}\}$ in all time slots. However, it is infeasible to perform 
the MAP decoding exactly in terms of the computational complexity. Instead, 
we derive suboptimal iterative decoders based on message-passing between a 
demodulator and a soft-input soft-output (SISO) decoder, using belief 
propagation (BP)~\cite{Pearl88,Richardson08} (See Fig.~\ref{fig2}). 
BP is a general algorithm for calculating marginal posterior probabilities 
for graphical models. If there are no cycles in the factor graph representing 
a graphical model, BP can calculate the marginal posterior probabilities 
exactly. BP may converge and provide a good approximation of the marginal 
posterior probabilities for a certain sparse factor graph, even if there are 
cycles in the factor graph. Notable examples are turbo 
codes~\cite{Berrou96,Kschischang98,McEliece98}, low-density parity-check 
(LDPC) codes~\cite{Richardson01}, multiuser 
decoding~\cite{Wang99,Boutros02,Caire04}, and iterative channel estimation and 
decoding~\cite{Alexander00,Valenti01,Vehkaperae091}. We believe that it is 
possible to show that BP-based iterative algorithms converge if the length of 
interleaving in BICM is sufficiently longer than the coherence 
time of the channels, by applying an argument in \cite{Richardson00}.  


\subsection{Soft-Decision Demodulator}
Existing BP-based SISO decoders can be used for calculating the messages from 
the SISO decoder to the demodulator in Fig.~\ref{fig2}. Thus, we only present 
the derivation of demodulators. 
The detection of (\ref{a_k}) is performed block by block. 
Without loss of generality, we focus on the first block of US, and drop 
the superscript from $a_{k}^{(j)}$. 
Let $P_{m}(x_{k,t})$ denote the message with respect to $x_{k,t}$ sent by 
the decoder in iteration~$m$. By the definition of 
BP~\cite[Chapter~2]{Richardson08}, the message $Q_{m+1}(x_{k,t})$ with 
respect to $x_{k,t}$ fed back to the decoder is given by 
\begin{equation}
Q_{m+1}(x_{k,t})\propto 
\sum_{a_{k}=0}^{1}p(a_{k})p(y_{k,t}|x_{k,t},a_{k})
\prod_{t'=0, t'\neq t}^{B-1}p(y_{k,t'}|a_{k}), \label{message} 
\end{equation}
with 
\begin{equation} \label{marginal_conditional_pdf} 
p(y_{k,t'}|a_{k}) = 
\sum_{\{x_{k,t'}\}}p(y_{k,t'}|x_{k,t'},a_{k}) P_{m}(x_{k,t'}). 
\end{equation}
In (\ref{message}), $p(a_{k})$ denotes a prior probability of (\ref{a_k}). 
Furthermore, the conditional probability density function 
(pdf) $p(y_{k,t'}|x_{k,t'},a_{k})$ represents the equivalent 
channel~(\ref{equivalent_channel}). In order to obtain an interpretable 
expression of (\ref{message}), we define the posterior probability of $a_{k}$ 
given $\{y_{k,t'}:t'=0,\ldots,B-1, t'\neq t\}$ as 
\begin{equation} \label{posterior} 
p(a_{k}|\{y_{k,t'}:t'\neq t\}) = \frac{
 p(a_{k})\prod_{t'\neq t}p(y_{k,t'}|a_{k})
}
{
 p(\{y_{k,t'}:t'\neq t\})
},  
\end{equation}  
with
\begin{equation}
p(\{y_{k,t'}:t'\neq t\}) = \sum_{a_{k}=0}^{1}p(a_{k})
\prod_{t'\neq t}p(y_{k,t'}|a_{k}). 
\end{equation}
Dividing the right-hand side of (\ref{message}) by the constant 
$p(\{y_{k,t'}:t'\neq t\})$ yields 
\begin{equation} \label{message_end} 
Q_{m+1}(x_{k,t})\propto 
\sum_{a_{k}=0}^{1}p(y_{k,t}|x_{k,t},a_{k})
p(a_{k}|\{y_{k,t'}:t'\neq t\}). 
\end{equation}
Since $a_{k}$ is a binary variable, the posterior 
probability~(\ref{posterior}) is characterized by the posterior mean 
$\hat{a}_{k}=\sum_{a_{k}=0}^{1}a_{k}p(a_{k}|\{y_{k,t'}:t'\neq t\})$. 

In order to evaluate the conditional pdf $p(y_{k,t}|x_{k,t},a_{k})$ we need 
the distribution of the interference $I_{k,t}$ in (\ref{equivalent_channel}). 
However, it is difficult to access its exact distribution. We use a Gaussian 
approximation instead: We approximate the distribution of $I_{k,t}$ by a 
circularly symmetric complex Gaussian distribution with variance 
$\sigma^{2}=\mathbb{E}[|I_{k,t}|^{2}]$. This approximation simplifies the 
conditional pdf $p(y_{k,t}|x_{k,t},a_{k})$, 
\begin{align} 
& p(y_{k,t}|x_{k,t},a_{k}) \nonumber \\  
\approx& \frac{a_{k}}{\pi N_{0}}\mathrm{e}^{
-\frac{|y_{k,t}-x_{k,t}/\sqrt{\mathcal{E}}|^{2}}{N_{0}}}
+ \frac{1-a_{k}}{\pi (N_{0}+\sigma^{2}/\mathcal{E})}
\mathrm{e}^{-\frac{|y_{k,t}|^{2}}{N_{0} + \sigma^{2}/\mathcal{E}}}. 
\label{conditional_pdf} 
\end{align} 

In calculating the message~(\ref{message_end}), we need the prior probability 
$p(a_{k})$, the noise variance $N_{0}$, the energy penalty, and 
the average power of the interference $\sigma^{2}$. For simplicity, 
we assume that the true values of these parameters are known in advance. 
In all numerical simulations, the true values are used. Note that it is 
straightforward to estimate these parameters in a decision-direct manner. 

In summary, the message $Q_{m+1}(x_{k,t})$ is updated as follows: 
The posterior probability~(\ref{posterior}), or equivalently the posterior 
mean $\hat{a}_{k}$, is first calculated 
from the prior probability $p(a_{k})$, (\ref{marginal_conditional_pdf}), 
(\ref{conditional_pdf}), and the messages $\{P_{m}(x_{k,t'})\}$.  
Next, the marginalization of the conditional pdf~(\ref{conditional_pdf})  
over $a_{k}$ is calculated to obtain the message~(\ref{message_end}). 
We refer to this demodulator as ``soft-decision demodulator.''  

\begin{remark}
We have assumed that perfect CSI is available at the transmitter. 
This assumption is an idealized assumption for TDD systems, in which the 
channel vectors are estimated on the basis of pilot signals 
transmitted through the reciprocal channel. If the channel estimation were 
imperfect, the equivalent channel~(\ref{equivalent_channel}) would include an 
additional interfering signal due to the channel estimation errors. There 
should not be much difference between the powers of the interfering signals 
for the data-dependent and data-independent schemes, if the channel estimation 
errors are independent of a subset of selected users $\mathcal{K}$. In other 
words, the interfering signals for both schemes should provide almost the same 
influence on the performance of the receiver. This argument allows us to 
assume perfect CSI at the transmitter, as long as the comparison between the 
data-dependent and data-independent schemes is concerned.  
\end{remark}

\subsection{Hard-Decision Demodulator}
In order to simplify the calculation of the message~(\ref{message_end}), 
we consider the hard decision of $a_{k}$ based on the MAP criterion
\begin{equation}
\hat{a}_{k}^{(\mathrm{MAP})} = \argmax_{a_{k}=\{0,1\}}
p(a_{k}|\{y_{k,t'}:t'\neq t\}). 
\end{equation} 
The message~(\ref{message_end}) is approximately calculated as 
\begin{equation} \label{message_MAP} 
Q_{m+1}^{(\mathrm{MAP})}(x_{k,t})\propto
p(y_{k,t}|x_{k,t},a_{k}=\hat{a}_{k}^{(\mathrm{MAP})}).  
\end{equation}
We refer to this demodulator as ``hard-decision demodulator.''  
The MAP detection of $a_{k}$ is equivalent to the maximum likelihood (ML) 
detection of $a_{k}$ for $\tilde{K}/K=1/2$. Note that the 
message~(\ref{message_MAP}) with respect to the data symbol $x_{k,t}$ 
is sent to the SISO decoder as soft information for both demodulators.  


\section{Numerical Simulations} \label{sec5} 
\subsection{Energy Efficiency}
The performance of the data-dependent US is numerically compared to that of the 
data-independent US, which is a greedy algorithm of US based on the 
criterion~(\ref{averaged_energy_penalty}), instead of 
(\ref{instantaneous_power}). As noted in Remark~\ref{remark3}, the 
data-independent US is a special case of greedy US proposed in \cite{Dimic05}. 
In all numerical simulations 
presented in this paper, quadrature phase shift keying (QPSK) is used. 
Furthermore, we assume independent and identically distributed (i.i.d.) 
Rayleigh block-fading channels with coherence time $T_{\mathrm{c}}$, i.e.\  
the channel vectors $\{\vec{\boldsymbol{h}}_{k,t}\}$ do not change during 
$T_{\mathrm{c}}$ time slots, and at the beginning of the next fading block 
they are independently sampled from a circularly symmetric complex Gaussian 
distribution with covariance matrix $\boldsymbol{I}_{N}$. 
A more practical assumption might be the assumption of block-fading with 
correlations between the adjacent blocks. However, the correlations provide 
no influence on the energy penalty~(\ref{average_energy_penalty}) if $B$ is 
smaller than the coherence time~$T_{\mathrm{c}}$, while they shorten the length 
of interleaving effectively. 

\begin{figure}[t]
\begin{center}
\includegraphics[width=\hsize]{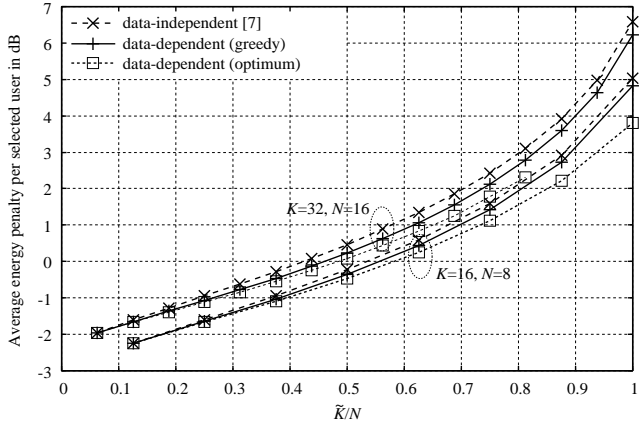}
\caption{
$\bar{\mathcal{E}}/\tilde{K}$ versus $\tilde{K}/N$ for $B=16$. 
}
\label{fig3} 
\end{center}
\end{figure}

We first focus on the performance of the data-dependent US algorithm 
in terms of the energy penalty (\ref{average_energy_penalty}) as 
$T\rightarrow\infty$. Figure~\ref{fig3} shows the average energy penalty per 
selected user with respect to $\tilde{K}/N$. For comparison, 
the optimal US based on the data-dependent 
criterion~(\ref{instantaneous_power}) and the (suboptimal) data-independent 
US are also plotted. The QPSK inputs 
were independently sampled with equal probability. This assumption is 
justified for proper error-correcting codes in conjunction with BICM. 
We find that the greedy algorithm of the data-dependent US outperforms the 
data-independent scheme, and that it can achieve nearly optimal energy 
penalty for small-to-moderate $\tilde{K}/N$.  

\begin{figure}[t]
\begin{center}
\includegraphics[width=\hsize]{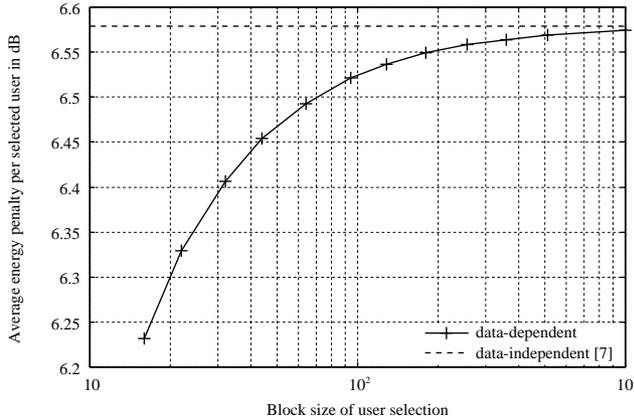}
\caption{
$\bar{\mathcal{E}}/\tilde{K}$ versus $B$ for $K=32$, $N=16$, and 
$\tilde{K}=16$. 
}
\label{fig4} 
\end{center}
\end{figure}

Figure~\ref{fig4} shows the average energy penalty versus the block size $B$ 
of US. The energy penalty for the data-dependent scheme increases slowly 
toward that for the data-independent scheme, as the block size $B$ grows. 
This observation is because it is unlikely that the amplitudes and 
phases of the interference $\vec{\boldsymbol{h}}_{k}
\boldsymbol{u}_{\mathcal{K}(i-1),t}$ 
in (\ref{energy_rec}) are close to those of the data symbols for 
all time slots. 

\subsection{BER} 
The bit error rate (BER) of the data-dependent scheme is compared to that of 
the data-independent scheme. It is preferable to use error-correcting codes 
satisfying the following three conditions: 
\begin{enumerate}
\item SISO decoding can be performed efficiently. 
\item High performance can be achieved in the low-rate regime. 
\item Robustness for erasures can be provided.  
\end{enumerate} 
Graph-based codes, such as turbo codes~\cite{Berrou96} and LDPC codes, 
satisfy the first condition. However, 
it is not straightforward to construct LDPC codes satisfying the second 
condition~\cite{Richardson04} (See also \cite{Richardson08}). 
For systematic codes such as turbo codes, the performance degrades 
significantly when the erasure of systematic bits occurs. Thus, a 
non-systematic code is a reasonable option for satisfying the last condition. 
In this paper, we use a repeat-accumulate (RA) 
code~\cite{Divsalar98,Abbasfar07} with rate~$r$. 
The RA code is a serial concatenation of a repetition code with 
rate $r$ and an accumulator. In BICM, QPSK is used in conjunction with random 
uniform interleaving whose length is equal to the code length $L/r$.  

\begin{figure}[t]
\begin{center}
\includegraphics[width=\hsize]{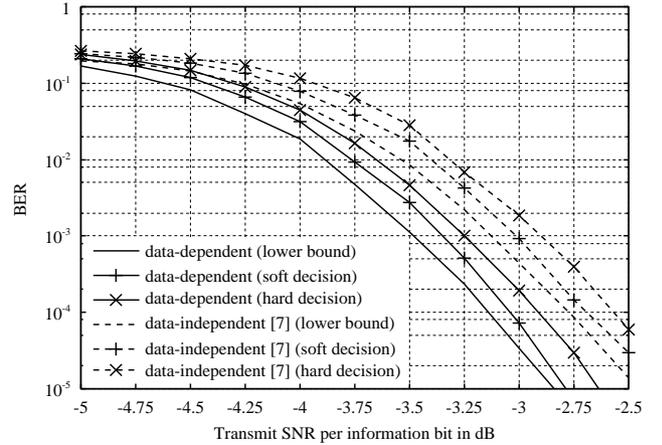}
\caption{
BER after 40 iterations. $K=32$, $\tilde{K}=16$, $N=16$,  
$B=16$, $T_{\mathrm{c}}=16$, $r=1/4$, and $L=4000$. 
}
\label{fig5} 
\end{center}
\end{figure}

Figure~\ref{fig5} presents the BERs for $N=B=T_{\mathrm{c}}=16$. The BERs of 
the data-dependent scheme for the soft-decision 
demodulator and the hard-decision demodulator are denoted by 
$\{+\}$ and $\{\times\}$ connected with solid lines, respectively. 
The messages are updated in the order ``demodulator 
$\rightarrow$ decoder for the inner code $\rightarrow$ decoder for the outer 
code $\rightarrow$ decoder for the inner code $\rightarrow$ 
demodulator $\rightarrow\cdots$''. The BER of genie-aided iterative decoding 
for the data-dependent scheme, in which a 
genie informs the receiver about (\ref{a_k}), is also shown by a 
solid line. Dashed lines are used, instead of solid lines, to represent the 
corresponding BERs for the data-independent schemes. The overall sum rate of 
all systems is equal to $2rK=16$~bps/Hz. The transmit SNR per information bit 
is defined as $1/(2rKN_{0})=1/(16N_{0})$. The data-dependent scheme can 
provide a performance gain of $0.35$~dB at a BER level 
of $10^{-4}$, compared to the data-independent scheme. 
The BERs of the soft-decision demodulator for both schemes are close to the 
corresponding genie-aided lower bounds. This implies that the soft-decision 
demodulator can detect successfully whether user~$k$ has been selected, i.e., 
(\ref{a_k}). The gaps between the soft-decision demodulator and the 
hard-decision demodulator correspond to the performance loss due to the hard 
decision of (\ref{a_k}). The soft decision of (\ref{a_k}) can achieve slightly 
smaller BER than that for the hard decision.  

\begin{figure}[t]
\begin{center}
\includegraphics[width=\hsize]{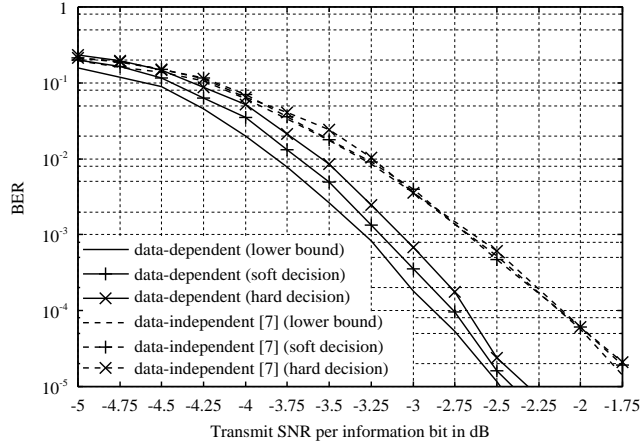}
\caption{
BER after 40 iterations. $K=32$, $\tilde{K}=16$, $N=16$, 
$B=16$, $T_{\mathrm{c}}=32$, $r=1/4$, and $L=4000$. 
}
\label{fig6} 
\end{center}
\end{figure}

Figure~\ref{fig6} shows the BERs for $N=16$ and $T_{\mathrm{c}}=32$. The block  
size of US for the data-dependent scheme is set to $B=16$\footnote{
When one makes a comparison between the proposed schemes in Figs.~\ref{fig5} 
and~\ref{fig6}, the comparison may be regarded as a comparison between 
users with different coherence times: Under the assumption that the 
coherence times are a multiple of $T_{\mathrm{c}}=16$, the proposed schemes 
in Fig.~\ref{fig5} show the performance for users with coherence 
time~$T_{\mathrm{c}}$, while those in Fig.~\ref{fig6} correspond to the case of 
users with coherence time~$2T_{\mathrm{c}}$.}, while the block size of 
US for the data-independent scheme is equal to the coherence time 
$T_{\mathrm{c}}=32$. Thus, the frequency of US for the data-dependent scheme is 
twice the frequency for the data-independent scheme. Interestingly, the 
diversity order (BER slope) for the data-dependent scheme is different from 
that for the data-independent scheme. This 
is because the block sizes of US for the two schemes are different from each 
other. The diversity order is determined by typical error events in the high 
SNR regime~\cite{Tse05}. The data symbols for non-selected users are erased 
during one block of US, i.e., during $B$ and $T_{\mathrm{c}}$ time slots for the 
data-dependent and data-independent schemes, respectively. The occurrence 
number of the erasure states around the mean fluctuates strongly as the block 
size of US increases. Decoding typically fails in the high SNR regime when the 
occurrence number of the erasure states deviates to a large value. As a 
result, the diversity order for the data-independent scheme is smaller than 
that for the data-dependent scheme.  

\begin{figure}[t]
\begin{center}
\includegraphics[width=\hsize]{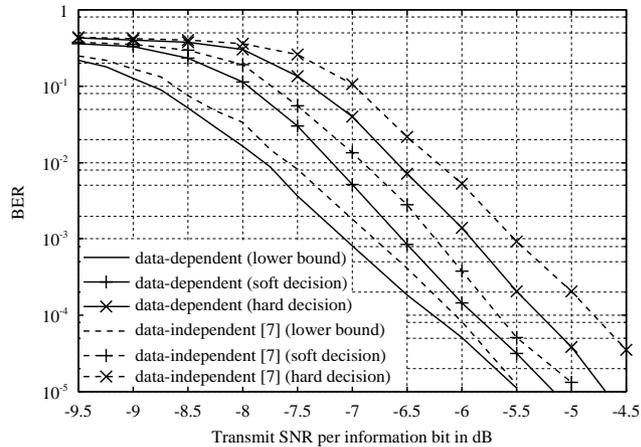}
\caption{
BER after 40 iterations. $K=32$, $\tilde{K}=8$, $N=16$, 
$B=16$, $T_{\mathrm{c}}=16$, $T_{\mathrm{c}}=16$, $r=1/8$, and $L=4000$. 
}
\label{fig7} 
\end{center}
\end{figure}

Figure~\ref{fig7} shows the BERs for $\tilde{K}=8$ and $K=32$. 
Since the RA code with $r=1/8$ is used, the overall sum rate 
of all systems is given by $8$~bps/Hz. The gaps between the genie-aided 
lower bounds and the BERs for the soft-decision demodulator are larger than 
those for $\tilde{K}=16$, shown in Fig.~\ref{fig5}. This observation is 
understood as follows: The code rate $r$ should be reduced as the ratio 
$\tilde{K}/K$ decreases. Reducing the rate $r$ results in decreasing a level 
of the receive SNR required for SISO decoding. Consequently, the demodulator 
is forced to detect (\ref{a_k}) for lower receive SNRs. This is the reason 
for the increase of the gaps between the lower bounds and the BERs for the 
soft-decision demodulator. Furthermore, we find that the gaps between the BERs 
for the two demodulators are also larger than those in Fig.~\ref{fig5}. 
This result implies that the soft decision of (\ref{a_k}) 
is an effective method for improving the performance for low SNRs. 

\subsection{Achievable Sum Rate} \label{sec53}
We have so far investigated the performance of the data-dependent scheme for 
fixed $\tilde{K}$ and $B$. How to choose $\tilde{K}$ and $B$ is discussed 
in this section. We first focus on $B$. One should choose $B\geq N$ in terms 
of the computational complexity, since the complexity per time slot is given 
by $O\{K\tilde{K}\max(1,N/B)N\}$, as shown in Section~\ref{sec_33}. 
The block size $B$ should be decreased in terms of the energy penalty, 
as shown in Fig.~\ref{fig4}, while $B$ should be increased in terms of the 
accurate detection of (\ref{a_k}).   
One reasonable option is to choose the smallest $B$ that achieves an accuracy 
requirement for the detection of (\ref{a_k}), determined by the used 
error-correcting codes.  

\begin{figure}[t]
\begin{center}
\includegraphics[width=\hsize]{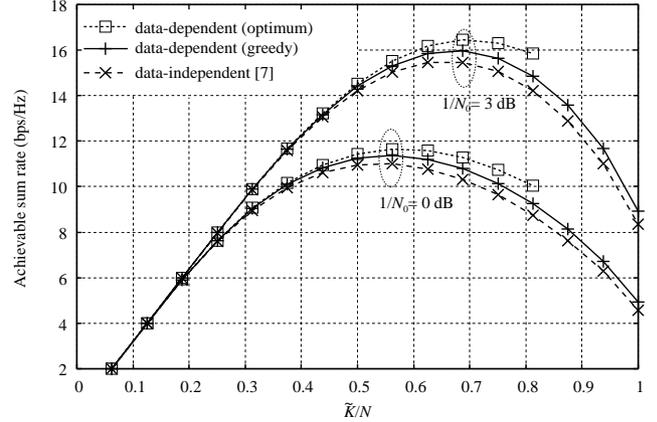}
\caption{
Achievable sum rare~(\ref{sum_rate}) 
versus $\tilde{K}$. $K=32$, $N=16$, and $B=16$. 
}
\label{fig8} 
\end{center}
\end{figure}

We next discuss how to choose $\tilde{K}$. We can assume that (\ref{a_k}) is 
known to the receiver, when $B$ is appropriately designed. Then, one should 
choose $\tilde{K}$ to maximize the achievable sum rate~(\ref{sum_rate}). 
Figure~\ref{fig8} plots the achievable sum rate~(\ref{sum_rate}) as a function 
of $\tilde{K}/N$. The achievable sum rates for the optimal data-dependent 
scheme and the data-independent scheme are also shown in the same figure. 
There are the optimal number $\tilde{K}_{\mathrm{opt}}$ of selected users for 
all SNRs $1/N_{0}$. The optimal number $\tilde{K}_{\mathrm{opt}}$ 
increases as SNR grows. These observations are consistent with 
the following information-theoretical intuitions: The whole power 
should be concentrated on sending messages for {\em one} user in the low 
SNR regime, while messages for multiple users should be sent simultaneously 
in the high SNR regime. The optimal number of selected users 
$\tilde{K}_{\mathrm{opt}}$ may be estimated in practice on the basis 
of feedback information about the receive SNR 
$1/(\bar{\mathcal{E}}N_{0})$, provided from each user.  

\section{Conclusions} \label{sec6} 
We have proposed a greedy algorithm of data-dependent US with ZFBF for fast 
fading Gaussian VBCs with perfect CSI at the transmitter. For the equal 
power case, the proposed US algorithm can outperform data-independent US 
in terms of the energy efficiency, BER, and the achievable sum rate, without 
increasing the complexity for the transmitter in terms of the order for fast 
fading channels. We have proposed iterative detection and decoding schemes 
based on BP. The schemes allow each user to detect whether he/she has been 
selected, without overhead for training, even for small block size of US. 
Furthermore, how to choose two design parameters has been discussed on the 
basis of the achievable sum rate. We conclude that data-dependent US is an 
efficient method of achieving a good tradeoff between the performance and 
the complexity for fast fading VBCs. 

\appendix

\section{Derivation of (\ref{user_rate})} 
\label{deriv_user_rate} 
Let us derive the lower bound~(\ref{user_rate}) on the achievable rate 
$R_{k}^{(\mathrm{opt})}$ of user~$k$ under the assumption that 
(\ref{energy_penalty}) and (\ref{a_k}) are known to the receiver. 
We know that the achievable rate $R_{k}^{(\mathrm{opt})}$ as 
$T\rightarrow\infty$ is equal to the mutual information per time slot between 
all data symbols and all variables known to the receiver~\cite{Tse05} 
\begin{equation}
R_{k}^{(\mathrm{opt})} 
= \lim_{T\rightarrow\infty}
\frac{1}{T}I(\{x_{k,t}\};\{y_{k,t}\},\{a_{k}^{(j)}\},\mathcal{E}), 
\end{equation}
which $y_{k,t}$ is given by (\ref{equivalent_channel}). 
Using the chain rule for mutual information~\cite{Cover06} yields  
\begin{align}
& R_{k}^{(\mathrm{opt})} \nonumber \\ 
=& \lim_{T\rightarrow\infty}\frac{1}{T}\left\{
 I(\{x_{k,t}\};\{a_{k}^{(j)}\},\mathcal{E})
 + I(\{x_{k,t}\};\{y_{k,t}\}|\{a_{k}^{(j)}\},\mathcal{E})
\right\} \nonumber \\ 
\geq& \lim_{T\rightarrow\infty}\frac{1}{T}
I(\{x_{k,t}\};\{y_{k,t}\}|\{a_{k}^{(j)}\},\mathcal{E})\;(\equiv R_{k}).
\label{lower_bound} 
\end{align}
In the derivation of the lower bound~(\ref{lower_bound}), we have used 
the non-negativity of mutual information. 
Since we are considering the assumption of fast fading, 
(\ref{energy_penalty}) is expected to converge in probability to a 
deterministic value $\bar{\mathcal{E}}$ as $T\rightarrow\infty$. Thus,  
(\ref{lower_bound}) reduces to 
\begin{equation}
R_{k} 
= \lim_{T\rightarrow\infty}\frac{1}{T}I(\{x_{k,t}\};\{y_{k,t}\}|\{a_{k}^{(j)}\}, 
\mathcal{E}=\bar{\mathcal{E}}). 
\end{equation} 

Let $\mathcal{J}_{\mathrm{s}} =\{j\in\{0,\ldots,T/B-1\}:a_{k}^{(j)}=1\}$ 
denote the set of the indices of blocks in which user~$k$ has been selected. 
We consider a suboptimal receiver that uses only the received signals 
in the blocks $j\in\mathcal{J}_{\mathrm{s}}$ to obtain a lower bound 
\begin{equation}
R_{k} 
\geq \lim_{T\rightarrow\infty}\frac{B\mathbb{E}[|\mathcal{J}_{\mathrm{s}}|]}{T}
C\left(
 \frac{1}{\bar{\mathcal{E}}N_{0}}
\right),  \label{lower_bound2}  
\end{equation}
with 
\begin{equation}
C\left(
\frac{1}{\bar{\mathcal{E}}N_{0}}
\right) = I(x_{k,t};y_{k,t}|a_{k}^{(j)}=1,\mathcal{E}=\bar{\mathcal{E}}),  
\end{equation}
which is given via the equivalent channel~(\ref{equivalent_channel}). 
In (\ref{lower_bound2}), the expectation of $|\mathcal{J}_{\mathrm{s}}|$ is 
defined as  
\begin{align}
&\mathbb{E}\left[
 |\mathcal{J}_{\mathrm{s}}|  
\right] 
= \sum_{\tilde{\mathcal{J}}_{\mathrm{s}}\subset\{0,\ldots,T/B-1\}} 
|\tilde{\mathcal{J}}_{\mathrm{s}}| \nonumber \\ 
&\times \mathrm{Prob}(\{a_{k}^{(j)}=1:j\in\tilde{\mathcal{J}}_{\mathrm{s}}\}, 
\{a_{k}^{(j)}=0:j\notin\tilde{\mathcal{J}}_{\mathrm{s}}\}).   
\end{align} 
The coefficient $\mathbb{E}[|\mathcal{J}_{\mathrm{s}}|]/(T/B)$ in 
(\ref{lower_bound2}) is 
equal to the average frequency at which user~$k$ is selected, and tends to 
$\tilde{K}/K$ in the limit $T\to\infty$, because of the assumption of fast 
fading. This implies that the lower bound~(\ref{lower_bound2}) reduces to 
(\ref{user_rate}).  

\section{Derivation of Data-Dependent US Algorithm} \label{appen1} 
We focus on the first block of US, and drop the superscript~$^{(j)}$ in 
(\ref{instantaneous_power}). 
The proposed US algorithm selects the user $\hat{k}$ to minimize 
(\ref{instantaneous_power}) with $\mathcal{K}=\mathcal{K}(i)$ in stage~$i$. 
We first prove that $E_{\mathcal{K}(i)}(B)$ is given by the recursive 
formula~(\ref{energy_rec}). Step~1 in the proposed algorithm implies that 
the statement holds for $i=1$. Thus, we assume $i>1$. Let us define 
$\boldsymbol{H}_{\mathcal{K}(i)}\in\mathbb{C}^{i\times N}$ as 
\begin{equation}
\boldsymbol{H}_{\mathcal{K}(i)} = 
\begin{bmatrix}
\boldsymbol{H}_{\mathcal{K}(i-1)} \\ 
\vec{\boldsymbol{h}}_{k}
\end{bmatrix}, 
\end{equation}
for $k\in\{1,\ldots,K\}\backslash\mathcal{K}(i-1)$. Substituting 
$\boldsymbol{u}_{\mathcal{K}(i),t}=\boldsymbol{H}_{\mathcal{K}(i)}^{\dagger}
\boldsymbol{x}_{\mathcal{K}(i),t}$ into (\ref{instantaneous_power}) with 
$\mathcal{K}=\mathcal{K}(i)$ yields 
\begin{equation} \label{instantaneous_power_tmp} 
E_{\mathcal{K}(i)}(B) = \frac{1}{B}\sum_{t=1}^{B}
\boldsymbol{x}_{\mathcal{K}(i),t}^{\mathrm{H}}
(\boldsymbol{H}_{\mathcal{K}(i)}
\boldsymbol{H}_{\mathcal{K}(i)}^{\mathrm{H}})^{-1}
\boldsymbol{x}_{\mathcal{K}(i),t}. 
\end{equation}
Using the inversion formula for block matrices, 
\begin{equation}
\begin{bmatrix}
\boldsymbol{A} & \boldsymbol{B} \\
\boldsymbol{C} & \boldsymbol{D} 
\end{bmatrix}^{-1}
= \begin{bmatrix}
\boldsymbol{A}^{-1}+\boldsymbol{A}^{-1}\boldsymbol{B}
\boldsymbol{E}^{-1}\boldsymbol{C}\boldsymbol{A}^{-1} & 
-\boldsymbol{A}^{-1}\boldsymbol{B}\boldsymbol{E}^{-1} \\ 
-\boldsymbol{E}^{-1}\boldsymbol{C}\boldsymbol{A}^{-1} & 
\boldsymbol{E}^{-1} 
\end{bmatrix}, 
\end{equation}
with $\boldsymbol{E}=\boldsymbol{D}-\boldsymbol{C}\boldsymbol{A}^{-1}
\boldsymbol{B}$, we obtain 
\begin{equation}
(\boldsymbol{H}_{\mathcal{K}(i)}
\boldsymbol{H}_{\mathcal{K}(i)}^{\mathrm{H}})^{-1} 
= \begin{bmatrix}
\boldsymbol{F} & 
-\frac{(\vec{\boldsymbol{h}}_{k}
\boldsymbol{H}_{\mathcal{K}(i-1)}^{\dagger})^{\mathrm{H}}}
{\vec{\boldsymbol{h}}_{k}\boldsymbol{P}_{i-1}^{\perp}
\vec{\boldsymbol{h}}_{k}^{\mathrm{H}}} \\ 
-\frac{\vec{\boldsymbol{h}}_{k}\boldsymbol{H}_{\mathcal{K}(i-1)}^{\dagger}}
{\vec{\boldsymbol{h}}_{k}\boldsymbol{P}_{i-1}^{\perp}
\vec{\boldsymbol{h}}_{k}^{\mathrm{H}}} & 
(\vec{\boldsymbol{h}}_{k}\boldsymbol{P}_{i-1}^{\perp}
\vec{\boldsymbol{h}}_{k}^{\mathrm{H}})^{-1}
\end{bmatrix}, \label{inverse} 
\end{equation}
with 
\begin{equation} \label{A} 
\boldsymbol{F}=(\boldsymbol{H}_{\mathcal{K}(i-1)}^{\dagger})^{\mathrm{H}}
\boldsymbol{H}_{\mathcal{K}(i-1)}^{\dagger} 
+ \frac{(\vec{\boldsymbol{h}}_{k}
\boldsymbol{H}_{\mathcal{K}(i-1)}^{\dagger})^{\mathrm{H}}
\vec{\boldsymbol{h}}_{k}\boldsymbol{H}_{\mathcal{K}(i-1)}^{\dagger}
}{\vec{\boldsymbol{h}}_{k}\boldsymbol{P}_{i-1}^{\perp}
\vec{\boldsymbol{h}}_{k}^{\mathrm{H}}}.
\end{equation} 
In (\ref{inverse}) and (\ref{A}), the Hermitian matrix 
$\boldsymbol{P}_{i-1}^{\perp}$ is the projection matrix~(\ref{P}) from 
$\mathbb{C}^{1\times N}$ onto the orthogonal complement 
of the subspace spanned by the channel vectors 
$\{\vec{\boldsymbol{h}}_{k}:k\in\mathcal{K}(i-1)\}$ 
selected in the preceding stages. Substituting the 
expression~(\ref{inverse}) into (\ref{instantaneous_power_tmp}) and 
using (\ref{instantaneous_power}) for $\mathcal{K}=\mathcal{K}(i-1)$, 
we arrive at the recursive formula~(\ref{energy_rec}).    

Next, we derive the recursive formula~(\ref{H}) for 
$\boldsymbol{H}_{\mathcal{K}(i)}^{\dagger}$. Expression~(\ref{H}) can be 
derived in the same manner as in the derivation of (\ref{energy_rec}): 
Substituting (\ref{inverse}) into 
$\boldsymbol{H}_{\mathcal{K}(i)}^{\dagger}= 
\boldsymbol{H}_{\mathcal{K}(i)}^{\mathrm{H}}
(\boldsymbol{H}_{\mathcal{K}(i)}
\boldsymbol{H}_{\mathcal{K}(i)}^{\mathrm{H}})^{-1}$, we immediately obtain 
the recursive formula~(\ref{H}).  

\bibliographystyle{ieicetr}
\bibliography{kt-ieice2011_2}






\end{document}